**Proton capture by $^{14}$N at astrophysical energies**

J. Grineviciute and Dean Halderson
Physics Department, Western Michigan University, Kalamazoo, MI 49008

**Abstract.** Calculations are reported for low energy, proton capture by $^{14}$N with the recoil corrected continuum shell model. An interaction from a fit to Cohen and Kurath (6-16) *p*-shell matrix elements and Reid soft core *g*-matrix elements is employed. The prediction for $^{14}$N($p,\gamma$)$^{15}$O, based on this model and available data, is that $S(0)$ equals 1.632 and S(30) is 1.625 keV b. Good agreement with available cross section data support this result. No evidence is found for significant contributions from the subthreshold resonance.

## 1. Introduction

Numerous nucleon capture reactions contribute to solar processes, however, for many the relevant energies are below those accessible in laboratories. Hence, extrapolation procedures have been employed to extend the measurements to lower energies. In [1] the recoil corrected continuum shell model (RCCSM) [2] was employed to extend measurements of the $^7$Be($p,\gamma$)$^8$B *S*-factor to the Gamow window and to zero energy.

The importance of the $^7$Be($p,\gamma$)$^8$B reaction in solar processes is well known. Many references discuss the need for accurate, low energy cross sections as a means of testing solar models and neutrino mixing. The significance of the $^{14}$N($p,\gamma$)$^{15}$O is its acting as a bottleneck reaction in the CNO cycle. Therefore, the reaction determines the energy production efficiency and CNO flux in the Sun.[3] By controlling the duration of hydrogen burning, it determines the main sequence turnoff, and thus ages of globular clusters.[4] Therefore the reaction has been studied extensively both experimentally and theoretically.

Theoretical work has included *R*-matrix fits [5] and a direct capture plus resonance calculation.[6] The latter is a single-particle calculation. A capture calculation for $^{14}$N($p,\gamma$)$^{15}$O based on a realistic Hamiltonian would be desirable to determine the behaviour of the *S*-factor down to zero energy and to test for consistency among data sets. However, this reaction presents a significant challenge to calculations from a realistic Hamiltonian. Whereas, $^8$B had only one state below proton threshold, $^{15}$O has seven to which the proton can capture. Since the important asymptotic behaviour of these states is determined by their energy from threshold, one would like an effective interaction that exactly reproduces all seven experimental energies as well as the $^1/_2{}^+$ resonance above threshold. It is unlikely that one can find an interaction which does this, and also gives reasonable fits to the other *p*-shell nuclei. An additional difficulty is the likelihood that $^{15}$O states contain components of great complexity, just as some $^{16}$O states have large components of 4*p*-4*h*.

However, the combination a calculation from a realistic Hamiltonian and available data can make accurate predictions for *S*-factors at low energies. The purpose of this investigation is to calculate the $S$-factors at low proton energies with a realistic interaction for each transition, and then to use those calculations and available data in the region 110 keV < $E_p$ < 350 keV where statistics are good to predict the $S$-factors below 110 keV. (All proton energies are centre of mass energies.) The sum of the $S$-factors can then be compare to a recent measurement of the sum of $S$-factors in the range 70 keV < $E_p$ < 110 keV. If consistency if achieved between the calculated and measured sum of $S$-factors in this energy range, then one has confidence in the prediction through the Gamow window and down to zero energy.

## 2. The Effective Interactions

The appropriate effective interaction depends on the model space employed. It has been found from calculations at the beginning of the $p$-shell that the best description of bound state structure and reactions is obtained when the core states that are nucleon emission stable are included in an RCCSM calculation. Therefore, the $1_1^+$, $0^+$, and $1_2^+$ states of $^{14}$N and the $0^+$ state of $^{14}$O in the $2h$ approximation are included in this calculation. An appropriate interaction for a continuum calculation requires an analytical form for the potential and not just the values for a set of matrix elements. The M3Y [7] interaction was shown to require modification as one moved toward the middle of the $p$-shell.[1] Calculations for $^8$B and $^8$Li favored a reduced tensor and increased spin-orbit interaction. In addition, a charge-symmetry breaking piece was necessary to account for the Nolan-Shiffer anomaly, and a Skyrme [8] component of the form, $V_S = t_3\delta(r_1 - r_2)\delta(r_2 - r_3)$, was necessary to obtain the correct thresholds. This form of the Skyrme interaction is required because the two-body, density dependent form is not translationally invariant.

In $^{14}$N and $^{15}$O the M3Y interaction has more difficulties, among which is to interchange the $0^+$ and $1_2^+$ levels in $^{14}$N. Therefore, the interaction mentioned in [1], which resulted from a fit to the Cohen and Kurath (6-16) $p$-shell matrix elements and Reid soft core $g$-matrix elements, is employed. However, the charge symmetry breaking interaction for $^8$B and $^8$Li and the Skyrme interaction were again required to obtain the correct thresholds. This combination provided a reasonable fit to the levels of $^8$B and $^8$Li. For $^{15}$O small adjustments in the spin-orbit and tensor strengths were required to match the positions of the $3/2_1^+$ state and the $1/2_2^+$ resonance. The spectrum for $^{15}$O is shown in figure 1 for a Skyrme strength of $t_3$ = 700 MeV fm$^6$, the CSB interaction of [11] multiplied by 0.555, and the interaction in the appendix labeled A. In order to provide a theoretical uncertainty, calculations are also made with a Skyrme strength of 1000 MeV fm$^6$ and the interaction in the appendix labeled B. A Skyrme strength any larger that 1000 MeV fm$^6$ brings the first $3/2^+$ state in the continuum down far too low. A characteristic of both interactions employed is to make the $5/2^+$ states come too low. Small adjustments in the interaction do not correct this.

One notices in figure 1 that the negative parity states come too low. The energy of the $^{3}/_{2}^{-}$ state in $^{15}$O is calculated to be 7 MeV below its observed energy. This is a typical result when calculations such as Hartree-Fock and the RCCSM, which determine single-particle wave functions that minimize the binding energies, do not use a density dependent interaction. Inner shells become too tightly bound. Density dependent interactions which depend on the coordinate $(\mathbf{r}_1 + \mathbf{r}_2)/2$ cannot be used in the RCCSM because they are not translationally invariant. The three-body Skyrme interaction given above has the effect of adding a density dependence. In fact, when $t_3$ is increased to 4500 MeV fm$^6$ and the central components of M3Y multiplied by 1.2, the bound state spectra of $^{14}$N and $^{15}$O, both positive and negative parity, look quite good. However, in the $^{15}$O continuum the $^{1}/_{2}{}^{+}_{2}$ and $^{3}/_{2}{}^{+}_{2}$ resonances are interchanged. Therefore, the positive parity and continuum states will be taken from either interactions A or B and the $^{1}/_{2}^{-}$ and $^{3}/_{2}^{-}$ wave functions taken from the modified M3Y plus Skyrme (MM3YS) interaction. This would seem to violate the spirit of the RCCSM where bound and continuum states are orthogonal because they come from the same Hamiltonian. However, both MM3YS and interactions A and B produce $^{1}/_{2}^{-}$ and $^{3}/_{2}^{-}$ states that are dominated by the $0\hbar\omega$, $p^{-1}$ configuration, and the main difference is the asymptotic behaviour due to the different energies. With the MM3YS, the $^{3}/_{2}^{-}$ wave function is 95.7% the $0\hbar\omega$, $p^{-1}_{3/2}$ state, while the wave function from interaction A is 98.2%.

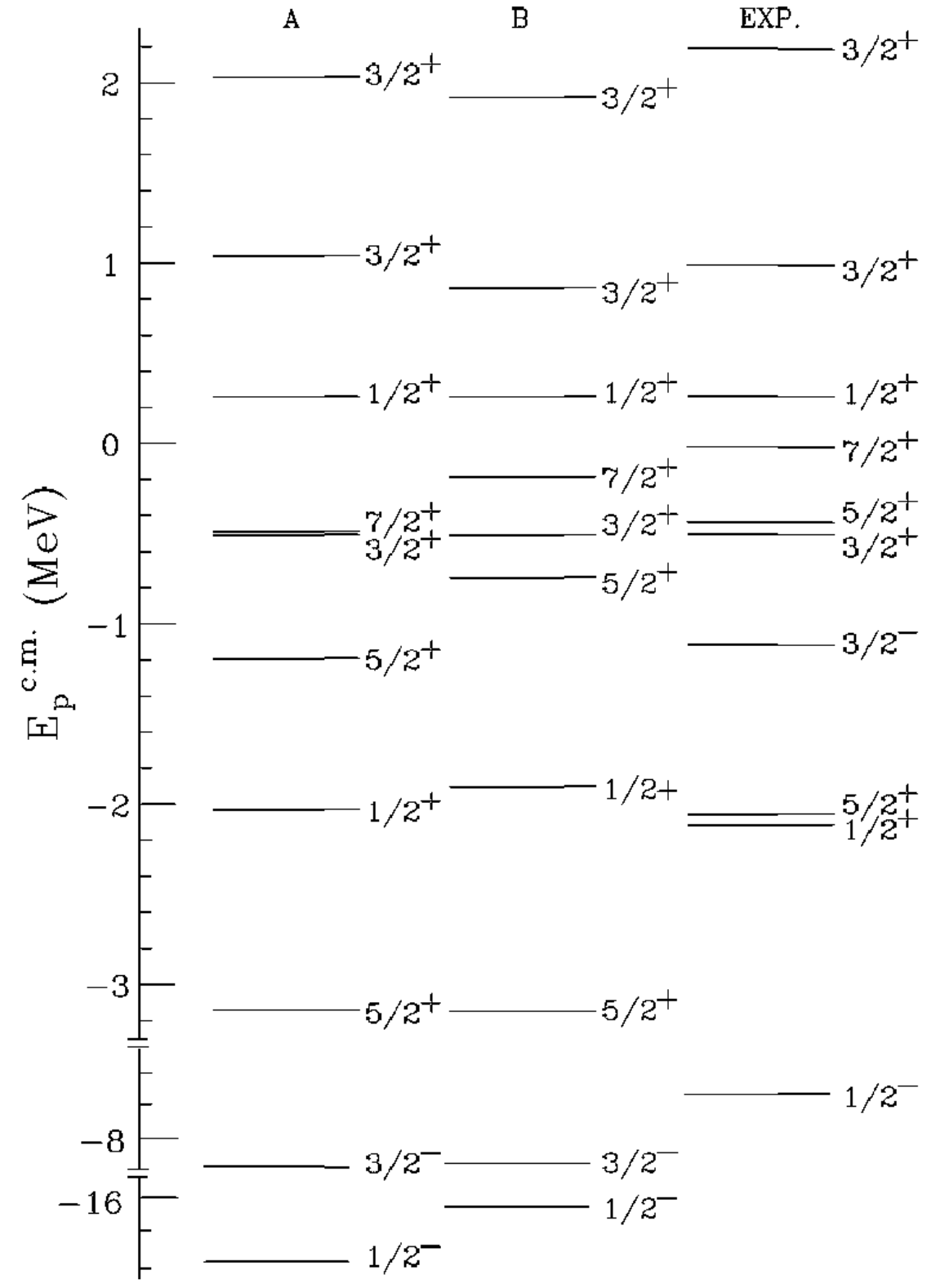

**Figure 1.** The $^{15}$O spectrum calculated with interactions A and B as compared to experiment

## 2. Procedure

The advantage of the RCCSM formalism is that it provides coupled-channels solutions for bound and unbound wave functions. The input to the RCCSM consists of only an oscillator size parameter, $\upsilon_0 = m\omega/\hbar$, the desired states of the $A-1$ core nuclei, and a realistic, translationally invariant interaction. An oscillator constant of $\upsilon_0 = 0.33$ fm$^{-2}$ is employed in this work. Since the calculations are transformed to the centre of mass system, the wave functions are antisymmetric and contain no spurious components. The procedure for calculating capture cross sections and analysing powers in a coupled-channels formalism is described in [12] and [13]. Wave functions and scattering matrices are calculated with $R$-matrix techniques.[14] This is not to be confused with an $R$-matrix fit where the energies and reduced widths are parameters as opposed to being calculated from a realistic Hamiltonian. The wave function with incoming flux $v_\alpha$ with target $\alpha$ has the form

$$\Psi_\alpha^{(+)} = \frac{4\pi}{k_\alpha} \sum_{lm_l jm J_B M_B} i^l Y^*_{lm_l}(\hat{k}_\alpha) e^{i\sigma_l} C^{l(1/2)j}_{m_l m_s m} C^{J_A j J_B}_{M_A m M_B} \psi^{J_B M_B}_{\alpha J_A lj} , \tag{1}$$

where

$$\psi^{J_B M_B}_{\alpha J_A lj} = \sum_{\alpha' J'_A l' j'} r^{-1}_{\alpha'} u^{J_B(+)}_{\alpha' J'_A l' j'}(r_{\alpha' J'_A l' j'}) |\alpha' J'_A l' j' J_B M_B\rangle , \tag{2}$$

and

$$u^{J_B(+)}_{c'}(r_{c'}) \to \frac{i}{2}\left(\frac{v_c}{v_{c'}}\right)^{1/2} (I_{c'}\delta_{c'c} - S_{c'c} O_{c'}) \tag{3}$$

outside the channel radius. The index $c$ stands for $ljJ_A\alpha J_B$; $v_c$ is a velocity, $I_c = G_c - iF_c$, $O_c = G_c + iF_c$, and $S_{c'c}$ is the $S$-matrix. For closed channels, the $S$-matrix element is zero and the radial wave function is proportional to a Whittaker function.

For $p$-shell nuclei [15] the channel wave functions within the channel radius, $a_c$, may be written as an expansion in an harmonic oscillator basis,

$$\psi^{J_B M_B}_{\alpha J_A lj} = \sum_{J_A \alpha \bar{l} \bar{j} \bar{n} \neq 0} f_{\bar{n}\bar{l}\bar{j} J_A \alpha J_B} \left[ a^+_{\bar{n}\bar{l}\bar{j}} \otimes |\alpha J_A\rangle \right]^{J_B} + \sum_\beta d_\beta |\beta J_B\rangle , \tag{4}$$

where $\beta$ runs over all $0\hbar\omega$, $p$-shell states with spin $J_B$, and $a^+_{\bar{n}\bar{l}\bar{j}}$ creates a particle in the core-nucleon, center of mass coordinate. The created particles are coupled to chosen $p$-shell states of the $A-1$ core, $|\alpha J_A\rangle$ . As mentioned above, the chosen core states are the $1^+_1$ , $0^+$, and $1^+_2$ states of $^{14}N$ and the $0^+$ state of $^{14}O$. The sum on $n$ cannot include zero when $l = 0$ or 1 because the $n = l = 0$ states are occupied and the $n = 0$, $l = 1$ states are included in the sum over $\beta$.

The differential capture cross section is given by

$$\frac{d\sigma}{d\Omega} = \frac{E_\gamma}{2\pi v_\alpha \hbar^2 c} \frac{1}{2[J_A]} \frac{\pi}{k^2_\alpha} \sum_{ljJ_BL\pi} \sum_{q=\pm 1} \sum_{l'j'J'_BL'\pi'} (i^l \hat{l} e^{i\sigma_l} T^{L\pi}_{J_b\alpha J_A ljJ_B})(i^{l'} \hat{l}' e^{i\sigma_{l'}} T^{L'\pi'}_{J_b\alpha J_A l'j'J'_B})^* \times$$
$$\sum_J \frac{[J_B][J'_B]}{\hat{l}} \hat{J}\hat{j}\hat{j}' (-1)^{j+j'-1-L'+q+J} C^{Jl'l}_{000} C^{LL'J}_{-qq0} P_J(\cos\theta) \times$$
$$W(JLJ'_BJ_b; L'J_B) W(JJ'_BjJ_A; J_Bj') W(Jj'l1/2; jl') , \tag{5}$$

where $\hat{x} = \sqrt{2x+1}$, $[x] = 2x + 1$, $J_b$ is the bound state spin,

$$T^{L\pi}_{J_b\alpha J_A l j J_B} = \langle \psi^{J_B}_{\alpha J_A l j} \| - T^{\pi}_{L} \| J_b \rangle^{*} , \tag{6}$$

and $T^{\pi}_{LM} = \alpha^{e}_{L}(Q_{LM} + Q'_{LM})$ or $q\alpha^{m}_{L}(M_{LM} + M'_{LM})$ for $(-1)^L = -1$ or $+1$. The effective multipole operators $Q_{LM}$ , $M_{LM}$, and $M'_{LM}$ are those given in Rose and Brink, [16] but modified to be translationally invariant.[1] Also, equation (6) is the only equation in this paper that uses the definition of the reduced matrix element in [16].

Therefore, one needs the reduced matrix elements of the electromagnetic operators between initial and final states. However, the *R*-matrix expansion in equation (4) is good within the channel radius, and, the transformation of matrix elements from shell model coordinates to the center of mass coordinates requires matrix elements of complete oscillators, not cut off at $a_c$. These matrix elements extend to infinity. Therefore a correction must be made to the transformed matrix elements in open channels where the wave functions extend beyond $a_c$. The contribution of the oscillator expansion beyond the channel radius must be subtracted, and the proper continuum wave function of equation (3) used beyond that point. Therefore, after the matrix elements have been transformed to the center of mass, they are first corrected by subtracting the contribution of the oscillator expansion outside of $a_c$. For example, the reduced matrix elements of the E1 operator, calculated with initial and final states in the form of equation (4), would receive the correction proportional to $\sum_{nn'} f^{*}_{nljJ_A\alpha J_B} f_{n'l'j'J'_A\alpha' J'_B} \int_{a_c}^{\infty} \langle \phi_{nlj} \| \sqrt{4\pi/3}\bar{e}Y_1(\hat{r})r \| \phi_{n'l'j'} \rangle \, dr$, where the oscillator wave functions, $\phi_{nlj}$, are calculated with a reduced $\upsilon = \upsilon_0(A-1)/A$. Outside the channel radius, the bound state channel wave functions becomes properly normalised Whittaker functions, $N_{c'} W_{-\eta_{c'},l+1/2}(2k_{c'}r)/r$, and the continuum state has a form given by equations (2) and (3). Hence, one adds contributions for each open channel proportional to $N_b$ $\int_{a_c}^{\infty} W_{-\eta_c,l+1/2}(2k_b r) r [I_c(k_c r) - O_{c'}(k_{c'} r) S_{cc'}] dr$ . At these low energies, the only open channels have $|\alpha J_A\rangle = {}^{14}\mathrm{N}(1^{+}_{1})$, however, all channels, $|\alpha' J'_A l' j' J_B M_B\rangle$ , contribute to the transition through equation (2).

The above procedure is demonstrated in figure 2 where the dominant imaginary piece of the integrand of the E1 operator beyond the channel radius, $\sum_{nn'} f^{*}_{nljJ_A\alpha J_B} f_{n'l'j'J'_A\alpha' J'_B} \langle \phi_{nlj} \| \sqrt{4\pi/3}\bar{e}Y_1(\hat{r})r \| \phi_{n'l'j'} \rangle$ , is plotted as if the expansion of equation (4) were extended beyond $a_c$, and also with the proper extension with Coulomb waves, $N_b$ $W_{-\eta_c,l+1/2}(2k_b r) r [I_c(k_c r) - O_{c'}(k_{c'} r) S_{cc'}]$. The curves are labeled with a I or II, respectively, to distinguish between the two cases. The plot is for $J_B = {}^{5}/_{2}^{-}$ channel $[{}^{14}\mathrm{N}(1^{+}_{1}) \otimes p_{3/2}]^{J_B=5/2}$ at $E_p = 0.485$ MeV connecting to the ${}^{3}/_{2}^{+}(6.791)$ closed channels, $[{}^{14}\mathrm{N}(1^{+}_{1}) \otimes s_{1/2}]^{J_B=3/2}$, $[{}^{14}\mathrm{N}(1^{+}_{1}) \otimes d_{3/2}]^{J_B=3/2}$, and $[{}^{14}\mathrm{N}(1^{+}_{1}) \otimes d_{5/2}]^{J_B=3/2}$ . One can see the integrand with the oscillator expansion and the integrand with the correct continuum form leaving from the channel radius with the same magnitude and slope. However, as shown in figure 3, the oscillators drop away quickly and the correct integrands extend to large radii. One also sees that the $[{}^{14}\mathrm{N}(1_1^+)\otimes s_{1/2}]^{J_B=3/2}$ channel is by far the dominant contributor from the final ${}^{3}/_{2}^{+}$(6.791) state. The imaginary part of the integrand for the channel to the dominant $^{3}/_{2}^{+}$ bound state. Solid line, dashed line, and dotted line are for $E_p = 0.485$ MeV, $5\times E_p = 0.259$ MeV, and $2000\times E_p = 0.070$ MeV.

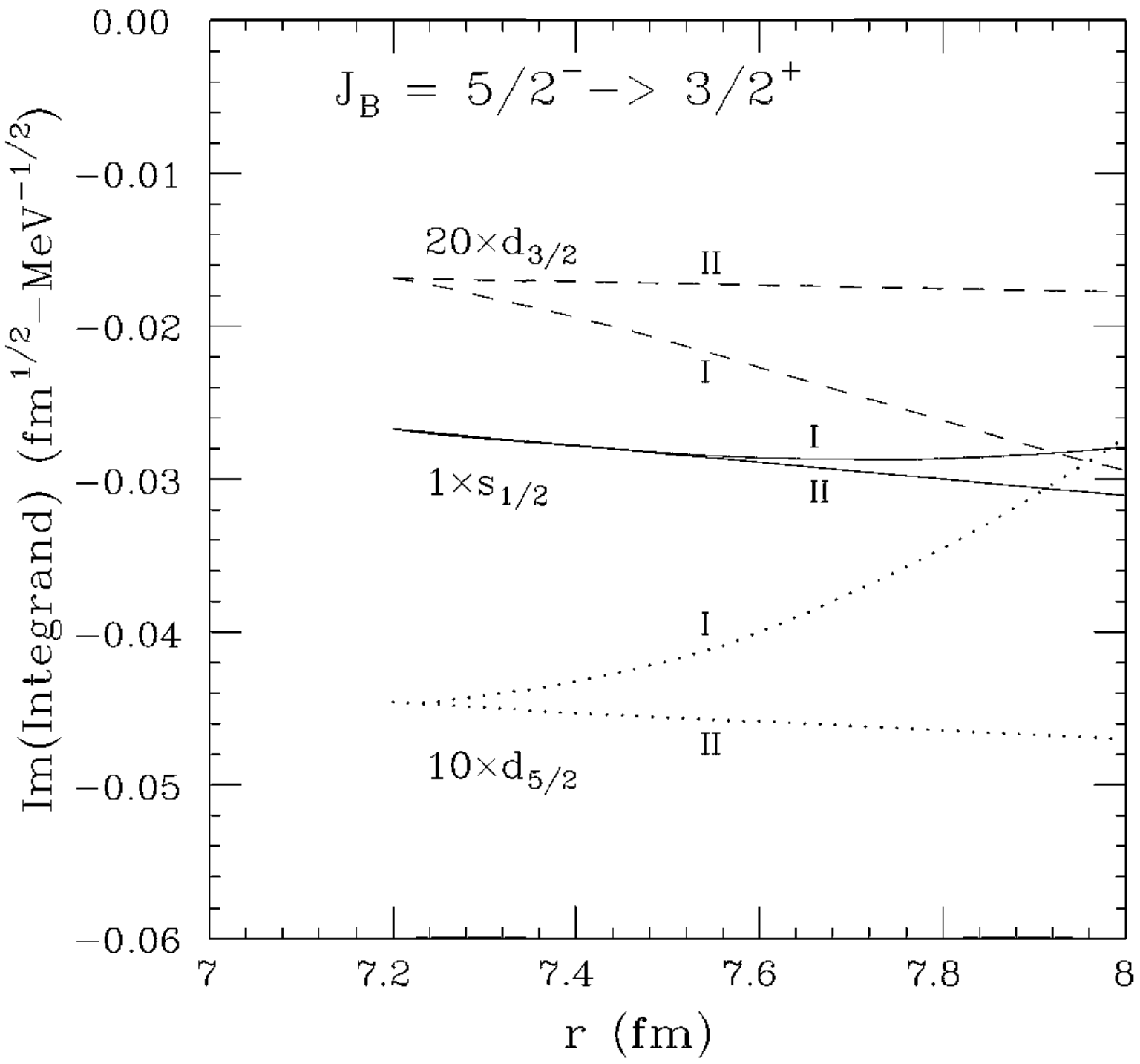

**Figure 2.** Imaginary parts of the radial integrands of the E1 operator at $E_p = 0.485$ MeV for the transition from the $J_B = {}^{5}/_{2}^{-}$ channel to the dominant ${}^{3}/_{2}^{+}$ bound state. Solid lines, dashed lines, and dotted lines are for the transition to the $s_{1/2}$, $20\times d_{3/2}$, and $10\times d_{5/2}$ component of the final state. Curves labeled I are for the oscillator expansion and II for the proper Coulomb waves.

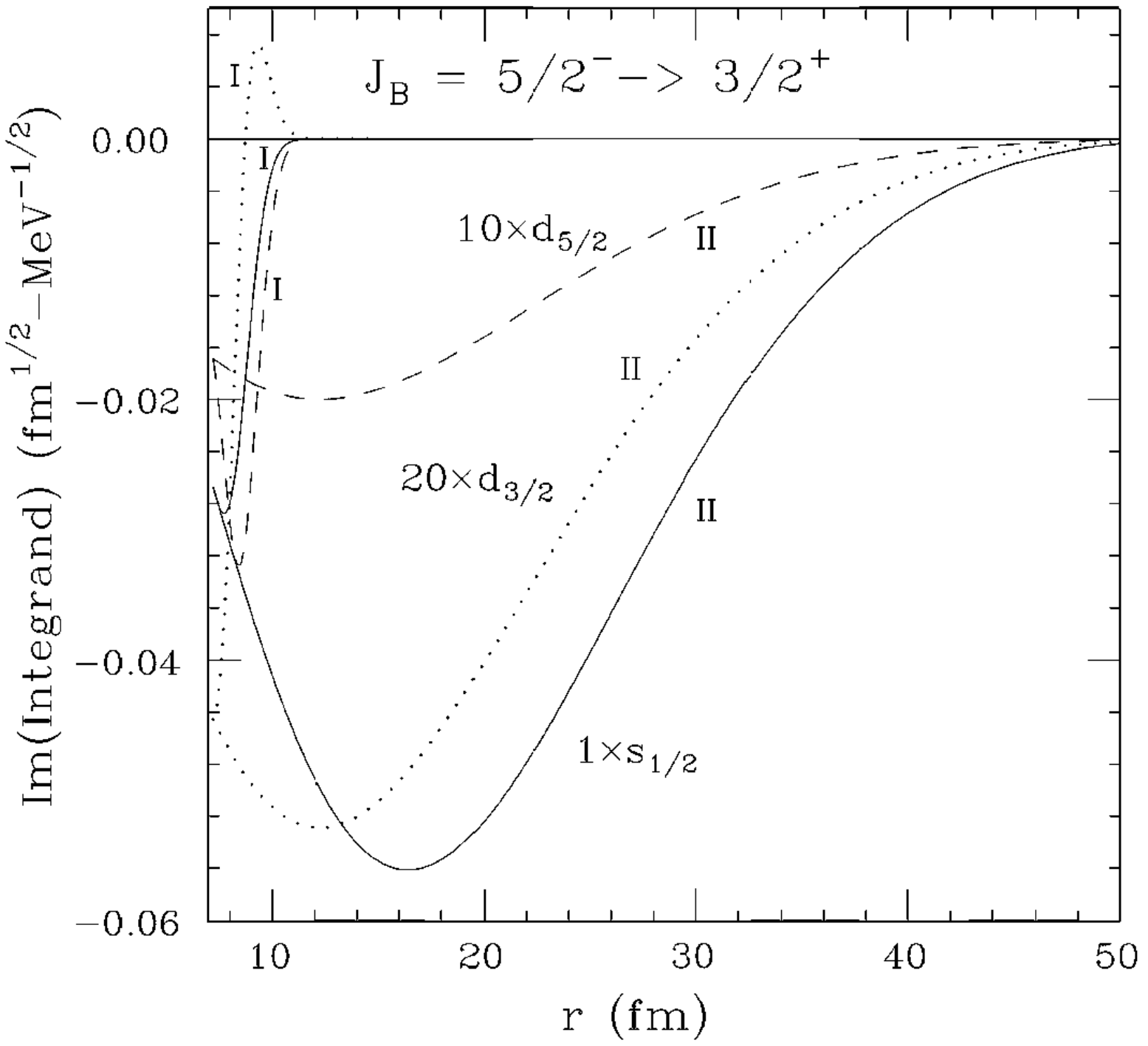

**Figure 3.** Same as figure 2

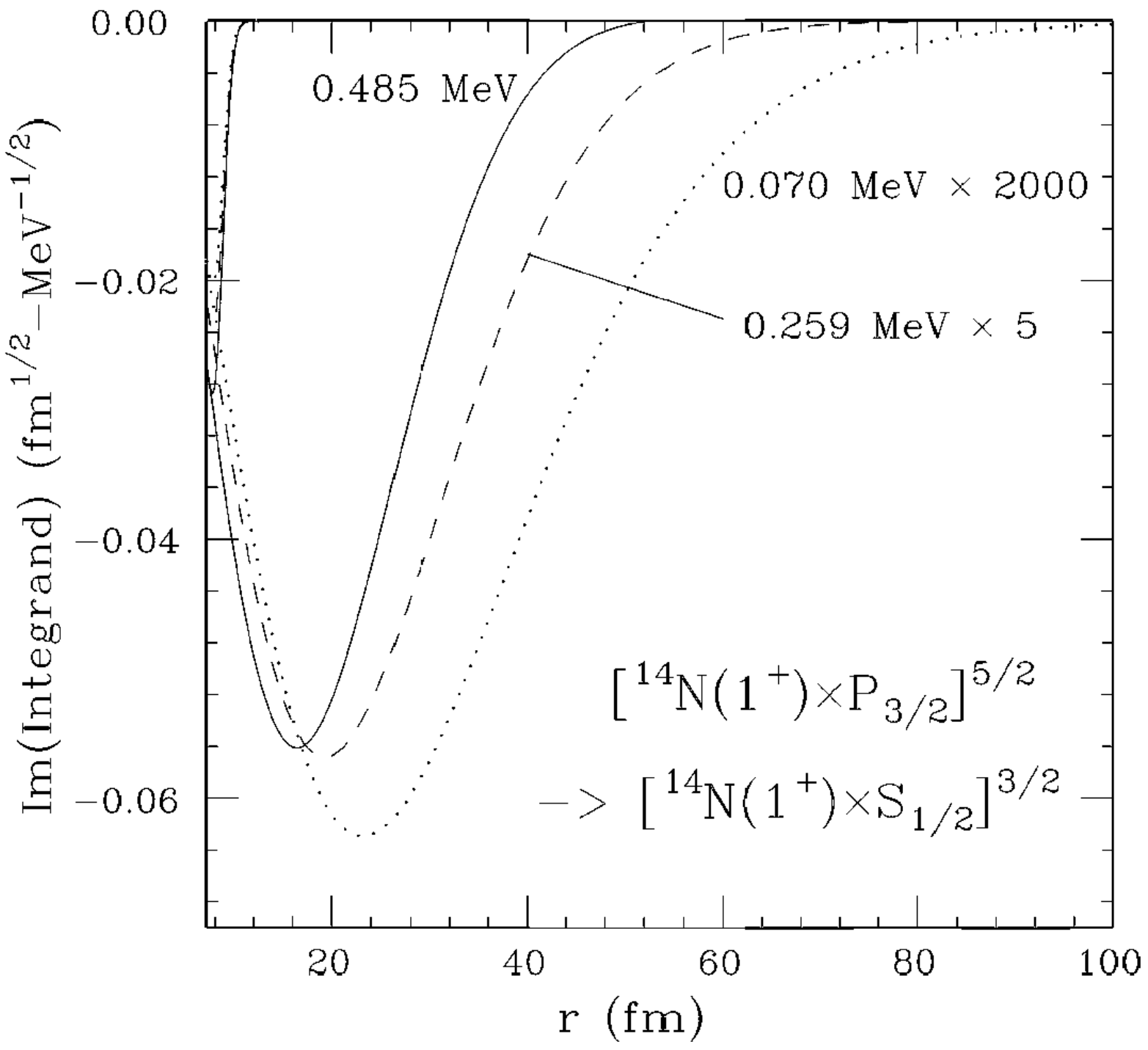

**Figure. 4.** Imaginary part of the integrands of the E1 operator for the $[^{14}\mathrm{N}(1_1^+) \otimes p_{3/2}]^{J_B=5/2} \rightarrow [^{14}\mathrm{N}(1_1^+) \otimes s_{1/2}]^{J_B=3/2}$ component of the transition from the $J_B = {}^5/_2{}^-$

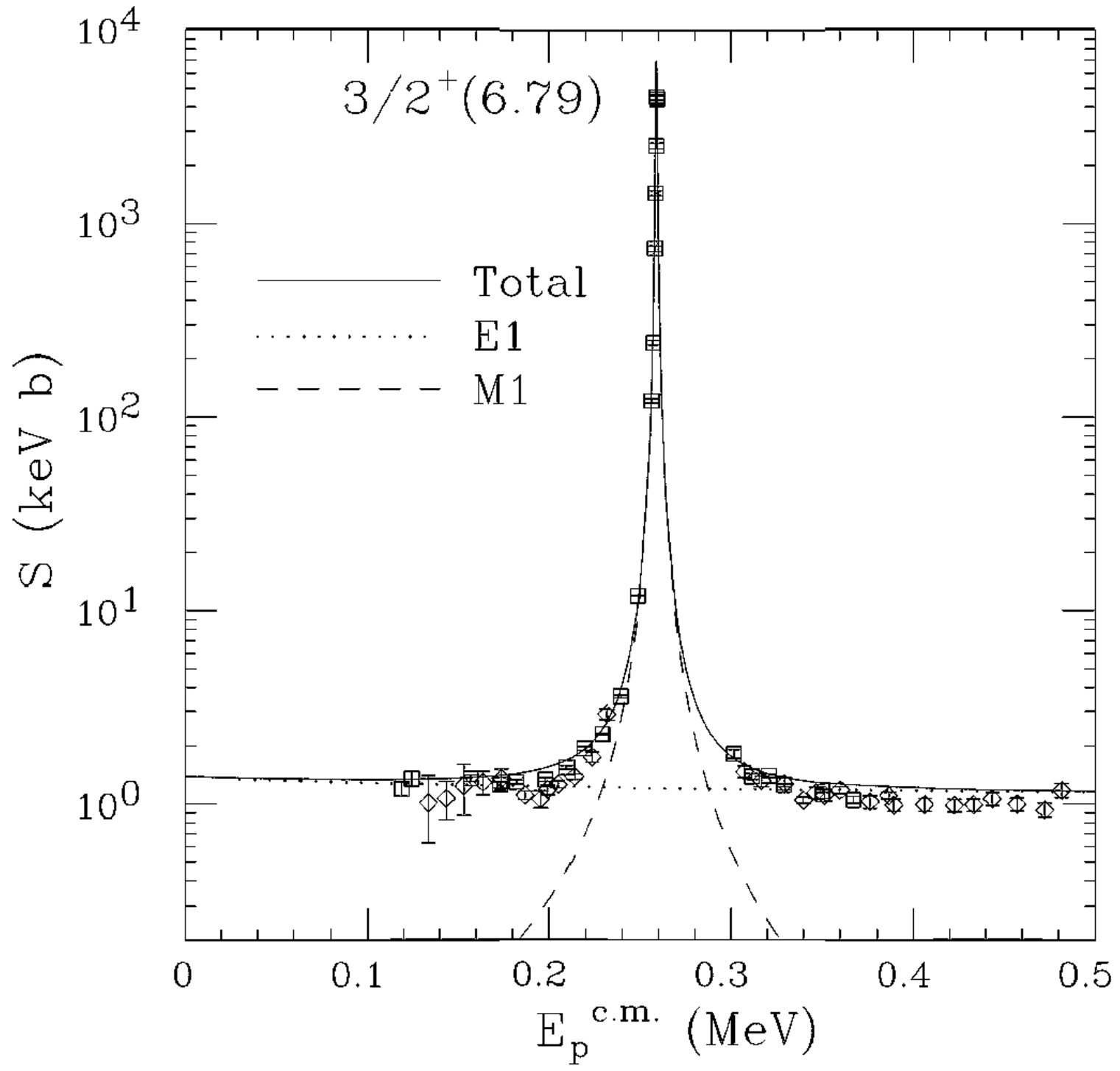

**Figure 5**. The *S*-factor for the transition to the $^{3}/_{2}^{+}$ (6.79 MeV) state in $^{15}O$. The squares are the data [17] diamonds of [18].

$[^{14}N(1_1^+)\otimes p_{3/2}]^{J_B=5/2} \rightarrow [^{14}N(1_1^+)\otimes s_{1/2}]^{J_B=3/2}$ transition is plotted in figure 4 for energies $E_p = 0.070$, 0.259, and 0.485 MeV to demonstrate the rapid decrease in size and the increase in radial extent.

## 3. The S-factors

The $^{14}N(p,\gamma)^{15}O$ *S*-factor at low energies is dominated by the transition to the $^{3}/_{2}^{+}$, 6.791 state. The results of the calculation for this transition with interaction A are shown in figure 5 along with the data of [17] as open squares and [18] as open diamonds. The calculation reproduces the width and M1 strength of the $^{1}/_{2}^{+}$ resonance very well. The E1 strength is slightly too large, and the E2 strength is negligible, not even appearing on the graph. One should note that the *S*-factor does not rise as the energy approaches zero as it did in $^{7}Be(p,\gamma)^{8}B$. A rise near zero energy is associated with very loosely bound systems, $^{17}F$ being the extreme example. An asymptotic normalization coefficient (ANC) of C = 4.5 and 4.6 $fm^{-1/2}$ was extracted from the best *R*-matrix fits in [18]. The calculated ANC for the dominant $[^{14}N(1_1^+)\otimes s_{1/2}]^{J_B=3/2}$ channel is 4.95 $fm^{-1/2}$. An angular distribution coefficient was extracted for this transition at $E_p = 0.485$ MeV in [19]. The value of $a_2 = -0.95$ is to be compared with the calculated value of -0.98. On the low energy side of the $^{1}/_{2}^{+}$ resonance, an angular distribution and analysing power were given in [6] at the effective proton energy of 0.245 MeV. The calculation at $E_p = 0.245$ MeV is shown in figure 5 as a solid line along with these data. By assuming an M1 resonance and E1 background, the authors of [6] obtained results similar to this calculation in a direct capture plus resonance model. Since the assumption of an E2 resonance gave results completely inconsistent with the analysing power, the authors of [6] concluded that the resonance strength must be M1. The present calculation certainly confirms this.

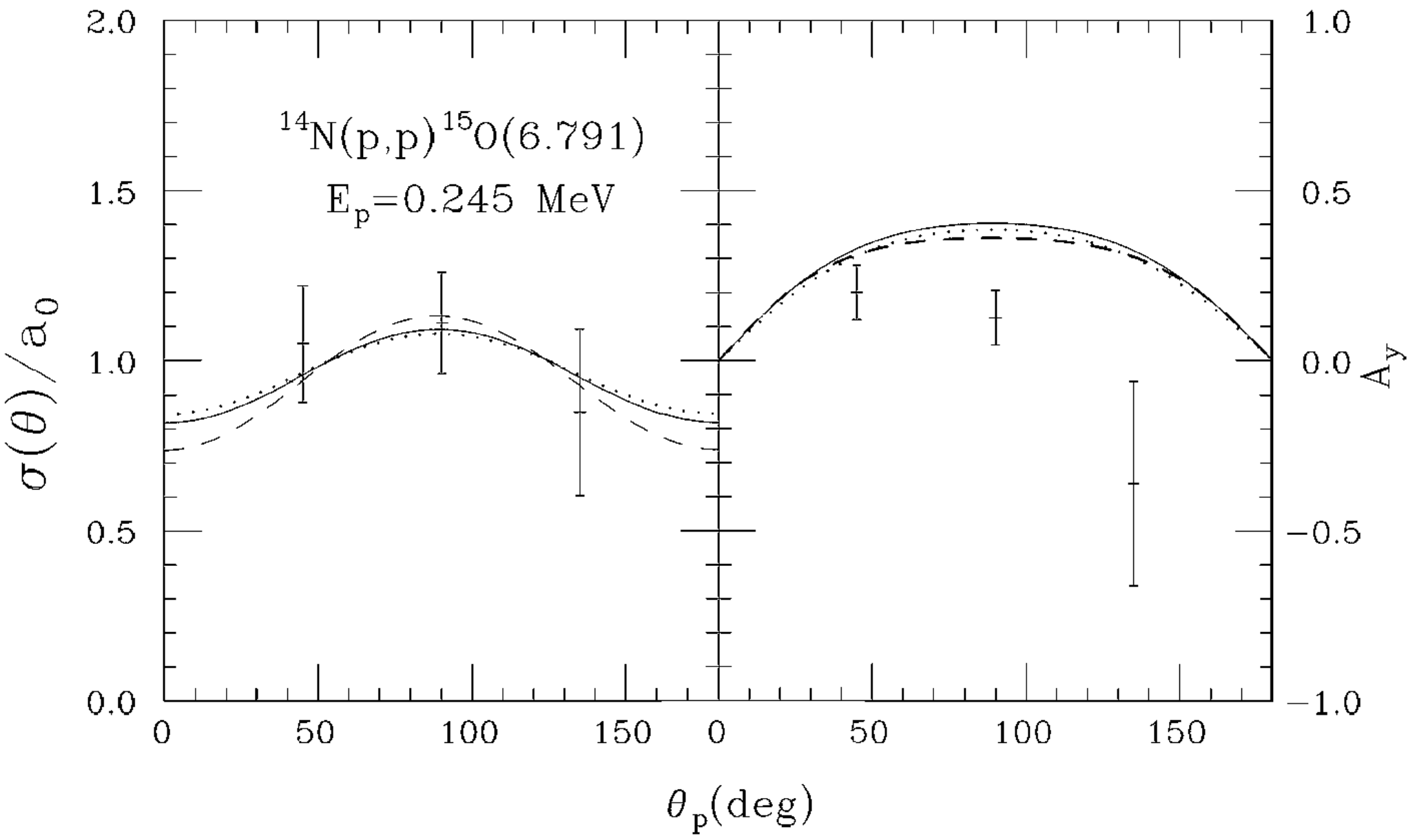

**Figure 6.** Angular distribution and analysing power for the transition to the $^{3}/_{2}^{+}$ (6.79 MeV) state in $^{15}O$. Data are from [6]. Solid line is from calculation at $E_p$ = 0.245 MeV. Dotted line is from calculation at $E_p$ = 0.245 MeV with scaled E1 background. Dashed line is calculation at effective energy, $E_p$ = 0.245 MeV, with scaled E1 background.

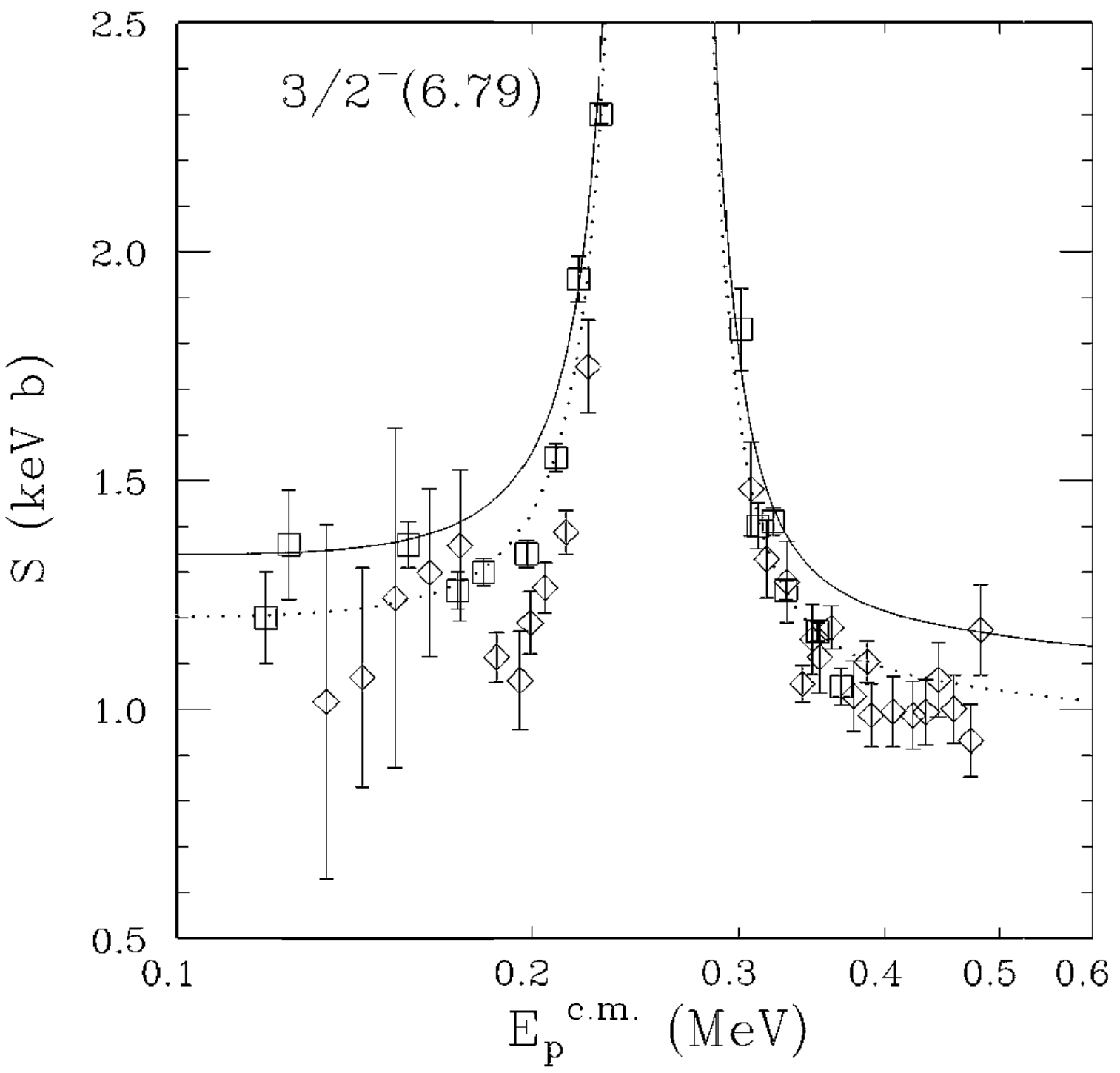

**Figure 7.** The *S*-factor for the transition to the $^{3}/_{2}^{+}$ (6.79 MeV) state in $^{15}O$. The squares are the data of [17], diamonds of [18]. The solid curve is the calculation; the dotted curve is the calculation with a scaled E1 background.

In order to extract an *S*-factor, the E1 background is scaled to minimize the $\chi^2$ with the data of [17] shown in figure 5 as open squares. This scaling of the E1 background was the same procedure used in [1] to extrapolate the $^{7}Be(p,\gamma)^{8}B$ data to zero energy. A scale factor of 0.895 gives the dotted curve in figure 7. This gives an *S*(30) of 1.221 keV b and an *S*(0) of 1.242 keV b. The calculation with interaction B required a scale factor of 1.131 and gives *S*(30)=1.208 keV b and *S*(0) = 1.242 keV b. The results are very similar because the backgrounds have similar energy dependences. However, this demonstrates that the magnitude of the background is sensitive to the interaction, and hence, the magnitude is difficult to predict.

When the E1 amplitude with interaction A is then scaled by $(0.895)^{1/2}$, the dotted curve in figure 6 is obtained. One sees very little change in the angular dependence. Second, the scaled calculation is made at an effective energy of 0.245 MeV by weighting the calculated cross sections from zero to 0.252 MeV by the total cross section divided by the stopping power for TiN [20] as was done for the direct capture plus resonance calculation of [6] The dashed line in figure 6 shows this result. Again one sees little change, especially in the analysing power.

The process of scaling the background contribution to minimize the $\chi^2$ with the data of [17] was repeated for the transition to the $^{5}/_{2}{}^{+}_{1}$ for interaction A. For the transition to the $^{1}/_{2}^{+}$ state, a better fit was obtained by scaling both the background and resonance. The background and resonance are both E1 for the transition to the $^{3}/_{2}^{-}$state. A summary of the scale factors is shown in table 1, and the resulting fits are shown as dotted lines in figure 8. Measurements in [21] found the contributions of the $^{5}/_{2}{}^{+}_{2}$, and $^{7}/_{2}^{+}$ to be negligible. The calculations confirm that the $^7/_2^+$ contribution is negligible, but find the $^5/_2{}^+_2$ state would contribute 0.022 keV b to $S(0)$ and and 0.023 keV b to $S(30)$.

As seen in table 1 and figure 8 the transition rate to the $^1/_2^-$ ground state is most unexpected for three reasons. First, the $^1/_2{}^+_2 \rightarrow {}^1/_2^-$ transition is an E1, and the rate should be large. Second, the gamma ray energy is 7.556 MeV, and the $k_\gamma^3$ dependence should make the rate large. Third, the $l = 1$ spectroscopic factor is $C^2S$=1.7 [22] which is 34 times that of the $^3/_2^-$ state. These factors would lead one to estimate the $S$-factor of the $^1/_2^-$ to be 1000 times that of the $^3/_2^-$. Indeed, that is what the calculation gives; whereas, the $^1/_2^-$ is measured to be an order of magnitude smaller. A possible explanation for this disagreement is found in the structure of the final state. If one considers the states in an intermediate coupling model, then the initial state is primarily $[^{14}\mathrm{N}(1_1^+)\otimes s_{1/2}]^{J_B=1/2}$, and the components of the ground state wave function which will connect to this can be written as $A[^{14}\mathrm{N}(1_1^+)\otimes p_{3/2}]^{J_B=1/2} + B[^{14}\mathrm{N}(1_1^+)\otimes p_{1/2}]^{J_B=1/2}$. The reduced matrix element of **r** between these initial and final states, $\langle f\|\mathbf{r}\|i\rangle$, would then be 0.272$A$ + 0.770$B$, where $A = -\ \langle^{15}\mathrm{O}(1/2^-)[a^+_{p_{1/2}}(p)\otimes|^{14}\mathrm{N}(1^+)\rangle]^{1/2}$ and $B = \langle^{15}\mathrm{O}(1/2^-)[a^+_{p_{3/2}}(p)\otimes|^{14}\mathrm{N}(1^+)\rangle]^{1/2}$ times an assumed common radial matrix element, $R$. The overlaps with interaction A are -1.036 and -0.460, respectively, giving $\langle f\|\mathbf{r}\|i\rangle$ = –0.072$R$. For Cohen and Kurath (6-16) the overlaps are –1.182 and –0.227, giving 0.147$R$. Therefore, the reduced matrix element is very sensitive to these overlaps, and one could certainly find interactions which give zero for the reduced matrix element. Stripping reactions have difficulty distinguishing between the $p_{1/2}$ and $p_{3/2}$ components, so they do not provide the sensitive information required for the overlaps. The correct $^{15}$O ground state would be much more complicated than this simple model; however, it must be some very fortunate cancellation such as that described above that produces the experimental result. The slightest variation would increase the hydrogen burning rate many times.

**Table 1.** Scale factors and primary multipoles for background and resonance contributions

| Final state | $^3/_2^+$ | $^3/_2^-$ | $^5/_2^+$ | $^1/_2^+$ | $^1/_2^-$(g.s.) |
|---|---|---|---|---|---|
| Primary background | E1 | E1 | M1 | E1 | E1($^3/_2^+$) |
| Primary resonance | M1 | E1 | E2 | M1 | E1($^1/_2^+$) |
| Background scale | 0.895 | 0.637 | 0.636 | 0.287 | 0.00909 |
| Resonance scale | 1. | 0.637 | 1. | 0.604 | 0.00016 |

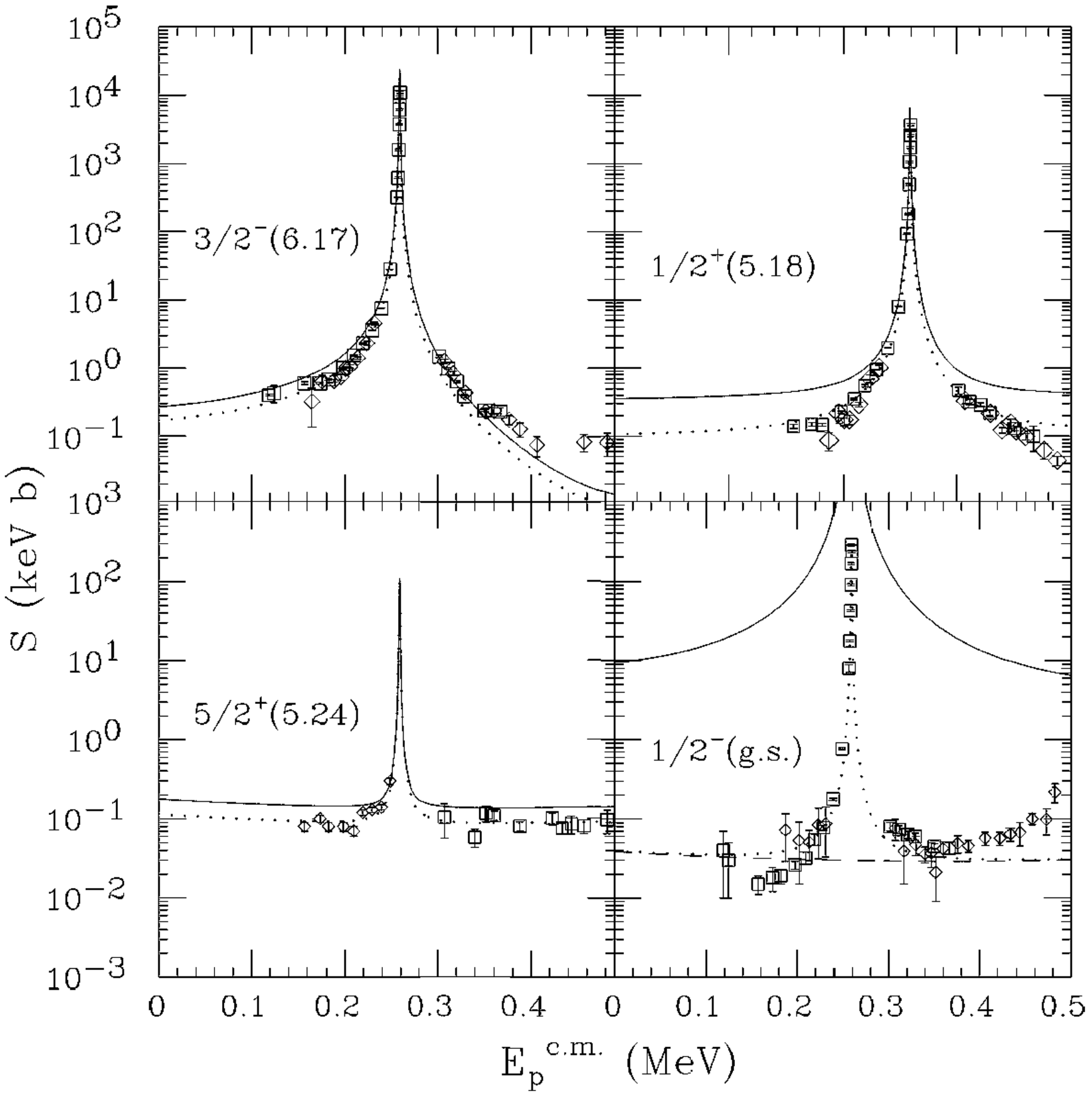

**Figure 8.** The $S$-factors for the transitions to the the $^3/_2^-$, $^1/_2^+$, $^5/_2{}^+_1$, and $^1/_2^-$ states in $^{15}$O. The squares are the data of [17], diamonds of [18]. The solid curve is the calculation; the dotted curve is the scaled calculation; and the dashed curve is the scaled $J_B = {}^3/_2^+$ continuum channel contribution to the transition to the ground state.

In contrast to the $^1/_2^-$, the coefficients for the $^3/_2^-$ state are $A = \langle ^{15}\mathrm{O}(1/2^-)[a^+_{p_{1/2}}(p)\otimes | {}^{14}\mathrm{N}(1^+)\rangle]^{3/2} = 0.326$ and $B = -\langle ^{15}\mathrm{O}(1/2^-)[a^+_{p_{3/2}}(p)\otimes | {}^{14}\mathrm{N}(1^+)\rangle]^{3/2} = -0.044$ for interaction A and 0.161 and –0.0003 for Cohen and Kurath (6-16), giving $\langle f \| \mathbf{r} \| i \rangle = 0.289R$ and 0.124R, respectively. Although the values differ by a factor of two, the reduced matrix element will not show the sensitivity found for the $^1/_2^-$ state. This is because the *B* coefficient is always small for realistic interactions and its sign is such that the two terms add coherently. Given the difference between these two values for $\langle f \| \mathbf{r} \| i \rangle$, the scale factor of 0.637 for interaction A seems a modest adjustment.

Because the $^1/_2^-$ state is sensitive to the interaction, a fit to the ground state capture data of [17] is performed by applying separate scale factors to the transition from the $J_B = {}^1/_2^+$ and $J_B = {}^3/_2^+$ continuum channels. The fit shown in figure 8 and is the poorest of the fits; however, the contribution to the total *S*-factor is small.

One should note that the *S*-factor for the $^1/_2^-$ ground state from the $J_B = {}^3/_2^+$ continuum channel, shown as a dashed line in figure 8, does not rise at low energies due to the subthreshold (6.791) state as it does in previous *R*-matrix fits. [5,19] In an RCCSM calculation the *R*-matrix level associated with this state contributes to the $J_B = {}^3/_2^+$ continuum channel through the amplitude,[23]

$$A_\mu = \frac{1}{E_\mu - E} \sum_{cc'} \gamma_{\mu d} b_c^0 (1 - R_B)^{-1}_{cc'} \gamma_{\mu_0 c'} A_{\mu_0}, \tag{7}$$

where $A_{\mu_0}$ is the amplitude of the level whose energy, $E_\mu$, is closest to $E$, $\gamma_{\mu c}$ is $(\frac{\hbar^2}{2m_c a_c})^{1/2} u_{\mu c}(a_c)$, $u_{\mu c}$ is a basis function in the channel $c$, $R_B$ is the *R*-matrix omitting the level closest to $E$, $b_c^0$ is $(\frac{r_c}{u_c^E}\frac{du_c^E}{dr_c})_{r_c = a_c}$, and $u_c^E(r_c)$ is the radial function of the physical wave function in the channel $c$. The $A_\mu$, in the notation of [23], corresponds to the $f_{\bar{n}\bar{l}\bar{j}J_A\alpha J_B}$ in equation 4 through a linear transformation. In an *R*-matrix fit, the contribution of this amplitude would be included in the parameters chosen for this level. In an RCCSM calculation, this amplitude depends only on the three inputs to the calculation: the oscillator constant, the chosen states of the core, and the interaction. The calculated amplitude is small for all interactions investigated, and the result is an *S*-factor which is flat through the Gamow window. Such a result is consistent with the analysis of [24] who find a simple pole in the *S*-factor at $E_\gamma = 0$. The *S*-factor is not expected to rise when $E_\gamma$ is large.

Finally, the sum of the scaled $^3/_2^+$, $^1/_2^+$, $^5/_{2\,1}^+$, $^1/_2^-$, and $^3/_2^-$ *S*-factors is plotted as a solid line in figure 9 with the corresponding data of [17] as open squares, [18] as open diamonds, and [21] as solid circles. The solid line goes through the squares because it was constructed by fitting the data of [17] which cover the110 keV < $E_p$ < 350 keV energy range. But the solid line also goes through the newer data of [21] that extends to down to 70 keV. This agreement provides confidence in the prediction of the calculation that $S(0)$ equals 1.676 and S(30) is 1.666 keV b. The solid curve is nearly flat for small energies. In [21] the authors assume that the $S$-factor remains constant below $E_p = 70$ keV in their calculation of the stellar reaction rates. The calculation provides justification for that assumption.

## IV. CONCLUSION

This paper has presented the RCCSM calculations for the reaction, $^{14}N(p,\gamma)^{15}O$. Calculations employed an interaction fit to a combination of the Cohen and Kurath (6-16) and Reid soft core matrix elements. Calculated angular distributions and $S$-factors for the transition to the dominant $^3/_2{}^+$ state agree well with available data. However, the calculated $S$-factor for the ground state transition is very different from that measured.

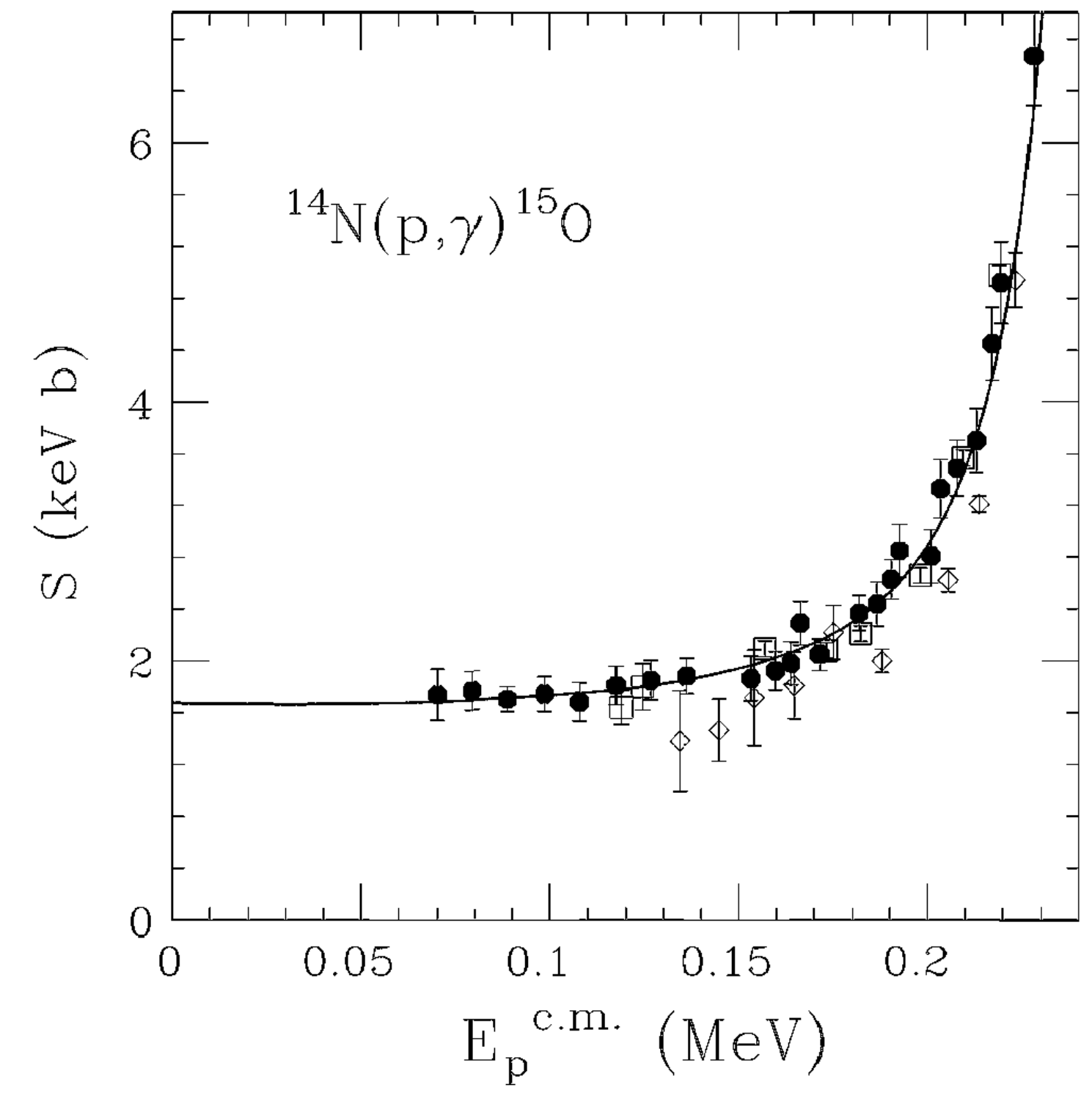

**Figure 9.** Sum of the scaled $^3/_2{}^+$, $^1/_2{}^+$, $^5/_2{}^+_1$, $^1/_2{}^-$, and $^3/_2{}^-$ $S$-factors from interaction A is plotted as a solid line. The solid circles are data of [21]; squares are the data of [17] and diamonds of [18] as given in [21].

An analysis of the ground state transition leads to the conclusion that an unexpected interference in the physical ground state wave function quenches the ground state transition. Without this quenching, the universe would not have evolved to its present state.

Calculated $S$-factors to individual transitions were scaled to fit the data of [17] in the110 keV $< E_p <$ 350 energy range. The sum of the scaled $^3/_2{}^+$, $^1/_2{}^+$, $^5/_2{}^+_1$, $^1/_2{}^-$, and $^3/_2{}^-$ $S$-factors gives $S(0)$ equals 1.676 keV b and $S(30) = 1.666$ keV b. The agreement between the energy dependence of the calculation and the data of [21] provides confidence in this result. The extension of the RCCSM calculation differs from published $R$-matrix fits in that the low energy *S*-factor is nearly flat. *R*-matrix fits allow an arbitrary contribution from the subthreshold $^{3}/_{2}^{+}$ state, and therefore have an *S*-factor that rises at low energies. The *R*-matrix level associated with the subthreshold $3/2^{+}$ state makes a small contribution to the RCCSM wave function, and hence, no rise in the *S*-factor.

## Acknowledgment

This work was supported by the National Science Foundation under grant PHY-0456943.

## References

[1] Halderson D 2006 *Phys. Rev.* C **73** 024612
[2] Philpott R J 1977 *Nucl. Phys.* A **289** 109
[3] Constantini H 2005 *Nucl. Phys.* A **758** 383c
[4] Imbriani G *et al.* 2005 *A. & A. J.* **420** 625
[5] Angulo C and Descouvemont P 2001 *Nucl. Phys.* A **690** 755
[6] Nelson S O *et al.*, 2003 *Phys. Rev.* C **68**, 065804
[7] Bertsch G, Borysowicz J, McManus H and Love W G 1977 *Nucl. Phys.* A **284** 399 (The interaction is described on p. 412.)
[8] Skyrme T H R 1959 *Nucl. Phys.* A **9** 615
[9] Cohen S and Kurath D 1965 *Nucl. Phys.* **73** 1
[10] Reid R V 1968 *Ann. Phys. NY* **50** 411
[11] Shlomo S 1978 S *Rep. Prog. Phys.* **41** 66
[12] Halderson D and Philpott R J 1979 *Phys. Rev. Lett.* **42** 36
[13] Halderson D and Philpott R J 1981 *Nucl Phys.* A **359** 365
[14] Philpott R J 1975 *Nucl. Phys.* A **243** 260
[15] Halderson D 2002 *Nucl. Phys.* A **707** 65
[16] Rose H J and Brink D M 1967 *Rev. Mod. Phys.* **39** 306
[17] Imbriani G *et al.*, 2005 *Eur. Phys. J* **A25** 455
[18] Runkle R C *et al.*, 2005 *Phys. Rev. Lett.* **94** 082503
[19] Schröder U, Becker H W, Bogaert B, Görres J, Rolfs C, Trautvetter H Pl, Azuma R E, Campbel C, King J D, and Vise J 1987 *Nucl. Phys.* A **467** 240
[20] Biersack J P and Zeigler J F 2006 program SRIM, version 2006.02
[21] Bemmerer D *et al.*, 2006 *Nucl. Phys.* A **779** 297
[22] Bertone P F, Champagne A E, Boswell M, Iliadis C, Hale S E, Hansper V Y and Powell D C 2002 *Phys. Rev.* C **66** 55804
[23] Philpott R J 1975 *Nucl. Phys.* A **243** 260
[24] Jennings B K, Karataglidis S, and Shoppa T D 1998 *Phys. Rev.* C **58** 3711
[25] Halderson D 1994 *J. Phys.* G **20** 1461

## Appendix

The effective interaction employs the same form and ranges as in [7] and [25]. The central components are given by $V = \sum_{i=1}^{3} V_i Y(r/R_i)$, the spin-orbit components by $V = \sum_{i=1}^{2} V_i Y(r/R_i)\, \boldsymbol{L}\cdot\boldsymbol{S}$, and the tensor components by $V = \sum_{i=1}^{2} V_i r^2 Y(r/R_i)\, S_{12}$, with $Y(x) = e^{-x}/x$. The coefficients, $V_i$, are given in Table 1 for interactions A and B.

**Table 2.** Effective interactions strengths, $V_i$.

| Int. | Force | Range (fm) | Triplet even (MeV) | Triplet odd (MeV) | Singlet even (MeV) | Singlet odd (MeV) |
|---|---|---|---|---|---|---|
| A | Central | 0.25 | 27120.71 | -800.000 | 6221.247 | 19999.04 |
| | | 0.40 | -8314.583 | 974.080 | -2338.178 | -1225.920 |
| | | 1.414 | -13.137 | 0. | -13.137 | 0. |
| | Spin-orbit | 0.25 | -910.000 | 9099.818 | | |
| | | 0.40 | -2728.726 | -2358.811 | | |
| | Tensor | 0.40 | -2256.124 | 582.981 | | |
| | | 0.70 | 71.925 | 16.439 | | |
| B | Central | 0.25 | 26783.75 | -732.745 | 6143.952 | 18317.75 |
| | | 0.40 | -8211.280 | 892.190 | -2309.128 | -1122.858 |
| | | 1.414 | -12.974 | 0. | -12.974 | 0. |
| | Spin-orbit | 0.25 | -840.000 | 8399.832 | | |
| | | 0.40 | -2518.824 | -2177.364 | | |
| | Tensor | 0.40 | -1869.2 | 483.000 | | |
| | | 0.70 | 59.590 | 13.620 | | |